\documentclass[aps,prl,reprint,superscriptaddress,floatfix,showpacs]{revtex4-1}
\usepackage[usenames,dvipsnames]{color}
\usepackage{natbib}
\usepackage{graphicx}
\usepackage{amsmath}
\usepackage{amssymb}
\usepackage{layouts}
\usepackage{tabularx}
\usepackage{array}
\usepackage{textcomp}
\usepackage{docs}
\usepackage{bm}

\usepackage[english]{babel}

\newcommand{\comment}[1]{ }

\begin{document}

\title{First Experimental Characterization of Microwave Emission
  from Cosmic Ray Air Showers}

\date{\today}

\author{R.~\v{S}m\'{\i}da}
\email[Corresponding author: ]{radomir.smida@kit.edu}
\affiliation{Karlsruhe Institute of Technology (KIT), Karlsruhe, Germany}
\author{F.~Werner}
\affiliation{Karlsruhe Institute of Technology (KIT), Karlsruhe, Germany}
\author{R.~Engel}
\affiliation{Karlsruhe Institute of Technology (KIT), Karlsruhe, Germany}
\author{J.C.~Arteaga-Vel\'azquez}
\affiliation{Universidad Michoacana, Instituto de F\'{\i}sica y Matem\'aticas, Morelia, M\'{e}xico}
\author{K.~Bekk}
\affiliation{Karlsruhe Institute of Technology (KIT), Karlsruhe, Germany}
\author{M.~Bertaina}
\affiliation{Universit\`{a} di Torino and Sezione INFN, Torino, Italy}
\author{J.~Bl\"{u}mer}
\affiliation{Karlsruhe Institute of Technology (KIT), Karlsruhe, Germany}
\author{H.~Bozdog}
\affiliation{Karlsruhe Institute of Technology (KIT), Karlsruhe, Germany}
\author{I.M.~Brancus}
\affiliation{National Institute of Physics and Nuclear Engineering, Bucharest, Romania}
\author{A.~Chiavassa}
\affiliation{Universit\`{a} di Torino and Sezione INFN, Torino, Italy}
\author{F.~Cossavella}
\altaffiliation[Present address: ]{DLR, Oberpfaffenhofen, Germany}
\affiliation{Karlsruhe Institute of Technology (KIT), Karlsruhe, Germany}
\author{F.~Di~Pierro}
\affiliation{Universit\`{a} di Torino and Sezione INFN, Torino, Italy}
\author{P.~Doll}
\affiliation{Karlsruhe Institute of Technology (KIT), Karlsruhe, Germany}
\author{B.~Fuchs}
\affiliation{Karlsruhe Institute of Technology (KIT), Karlsruhe, Germany}
\author{D.~Fuhrmann}
\altaffiliation[Present address: ]{University of Duisburg-Essen, Duisburg, Germany}
\affiliation{Bergische Universit\"{a}t Wuppertal, Wuppertal, Germany}
\author{C.~Grupen}
\affiliation{Department of Physics, Siegen University, Germany}
\author{A.~Haungs}
\affiliation{Karlsruhe Institute of Technology (KIT), Karlsruhe, Germany}
\author{D.~Heck}
\affiliation{Karlsruhe Institute of Technology (KIT), Karlsruhe, Germany}
\author{J.R.~H\"orandel}
\affiliation{Department of Astrophysics, Radboud University Nijmegen, The Netherlands}
\author{D.~Huber}
\affiliation{Karlsruhe Institute of Technology (KIT), Karlsruhe, Germany}
\author{T.~Huege}
\affiliation{Karlsruhe Institute of Technology (KIT), Karlsruhe, Germany}
\author{K.-H.~Kampert}
\affiliation{Bergische Universit\"{a}t Wuppertal, Wuppertal, Germany}
\author{D.~Kang}
\affiliation{Karlsruhe Institute of Technology (KIT), Karlsruhe, Germany}
\author{H.~Klages}
\affiliation{Karlsruhe Institute of Technology (KIT), Karlsruhe, Germany}
\author{M.~Kleifges}
\affiliation{Karlsruhe Institute of Technology (KIT), Karlsruhe, Germany}
\author{O.~Kr\"{o}mer}
\affiliation{Karlsruhe Institute of Technology (KIT), Karlsruhe, Germany}
\author{K.~Link}
\affiliation{Karlsruhe Institute of Technology (KIT), Karlsruhe, Germany}
\author{P.~{\L}uczak}
\affiliation{National Centre for Nuclear Research, Department of Astrophysics, {\L}\'{o}d\'{z}, Poland}
\author{M.~Ludwig}
\affiliation{Karlsruhe Institute of Technology (KIT), Karlsruhe, Germany}
\author{H.J.~Mathes}
\affiliation{Karlsruhe Institute of Technology (KIT), Karlsruhe, Germany}
\author{S.~Mathys}
\affiliation{Bergische Universit\"{a}t Wuppertal, Wuppertal, Germany}
\author{H.J.~Mayer}
\affiliation{Karlsruhe Institute of Technology (KIT), Karlsruhe, Germany}
\author{M.~Melissas}
\affiliation{Karlsruhe Institute of Technology (KIT), Karlsruhe, Germany}
\author{C.~Morello}
\affiliation{Osservatorio Astrofisico di Torino, INAF Torino, Italy}
\author{P.~Neunteufel}
\affiliation{Karlsruhe Institute of Technology (KIT), Karlsruhe, Germany}
\author{J.~Oehlschl\"ager}
\affiliation{Karlsruhe Institute of Technology (KIT), Karlsruhe, Germany}
\author{N.~Palmieri}
\affiliation{Karlsruhe Institute of Technology (KIT), Karlsruhe, Germany}
\author{J.~Pekala}
\affiliation{Institute of Nuclear Physics PAN, Krakow, Poland}
\author{T.~Pierog}
\affiliation{Karlsruhe Institute of Technology (KIT), Karlsruhe, Germany}
\author{J.~Rautenberg}
\affiliation{Bergische Universit\"{a}t Wuppertal, Wuppertal, Germany}
\author{H.~Rebel}
\affiliation{Karlsruhe Institute of Technology (KIT), Karlsruhe, Germany}
\author{M.~Riegel}
\affiliation{Karlsruhe Institute of Technology (KIT), Karlsruhe, Germany}
\author{M.~Roth}
\affiliation{Karlsruhe Institute of Technology (KIT), Karlsruhe, Germany}
\author{F.~Salamida}
\altaffiliation[Present address: ]{Institut Physique Nucl\'{e}aire d'Orsay, Orsay, France}
\affiliation{Karlsruhe Institute of Technology (KIT), Karlsruhe, Germany}
\author{H.~Schieler}
\affiliation{Karlsruhe Institute of Technology (KIT), Karlsruhe, Germany}
\author{S.~Schoo}
\affiliation{Karlsruhe Institute of Technology (KIT), Karlsruhe, Germany}
\author{F.G.~Schr\"oder}
\affiliation{Karlsruhe Institute of Technology (KIT), Karlsruhe, Germany}
\author{O.~Sima}
\affiliation{Department of Physics, University of Bucharest, Bucharest, Romania}
\author{J.~Stasielak}
\affiliation{Institute of Nuclear Physics PAN, Krakow, Poland}
\author{G.~Toma}
\affiliation{National Institute of Physics and Nuclear Engineering, Bucharest, Romania}
\author{G.C.~Trinchero}
\affiliation{Osservatorio Astrofisico di Torino, INAF Torino, Italy}
\author{M.~Unger}
\affiliation{Karlsruhe Institute of Technology (KIT), Karlsruhe, Germany}
\author{M.~Weber}
\affiliation{Karlsruhe Institute of Technology (KIT), Karlsruhe, Germany}
\author{A.~Weindl}
\affiliation{Karlsruhe Institute of Technology (KIT), Karlsruhe, Germany}
\author{H.~Wilczy\'{n}ski}
\affiliation{Institute of Nuclear Physics PAN, Krakow, Poland}
\author{M.~Will}
\affiliation{Karlsruhe Institute of Technology (KIT), Karlsruhe, Germany}
\author{J.~Wochele}
\affiliation{Karlsruhe Institute of Technology (KIT), Karlsruhe, Germany}
\author{J.~Zabierowski}
\affiliation{National Centre for Nuclear Research, Department of Astrophysics, {\L}\'{o}d\'{z}, Poland}

\begin{abstract}
We report the first direct measurement of the overall characteristics of
microwave radio emission from extensive air showers. Using a trigger
provided by the KASCADE-Grande air shower array, the signals of the
microwave antennas of the CROME (Cosmic-Ray Observation via Microwave
Emission) experiment have been read out and searched for signatures of
radio emission by high-energy air showers in the GHz frequency range.
Microwave signals have been
detected for more than 30 showers with energies above
$3\times10^{16}$\,eV. The observations presented in this Letter are consistent
with a mainly forward-directed and polarised emission process in the GHz
frequency range.
The measurements show that microwave radiation offers a new means of studying
air showers at $E\geq10^{17}$\,eV.
\end{abstract}


\maketitle


\textit{Introduction} --
At energies above $10^{15}$\,eV cosmic rays can be measured only indirectly by studying the
extensive air showers they produce in the Earth's atmosphere~\cite{Bluemer:2009zf}.
Different techniques have been developed to measure air showers with instruments of large
aperture. While arrays of particle detectors at ground can only sample the shower at
one particular depth in the atmosphere, optical detectors allow the measurement of the evolution
of the shower, including the depth profile of the number of shower particles in the atmosphere.
Optical methods have the advantage of providing a calorimetric measurement of the shower
energy and, through determining the depth of shower maximum, also a good estimate of the
type or mass of the primary particle. On the other hand, they can be applied only in dark nights
and good atmospheric conditions, limiting the duty cycle to typically less than 15\%.

Since the pioneering studies in the late
1960s~\cite{Allan:1971aa} it has been known that extensive air showers
produce electromagnetic pulses in the kHz and MHz frequency range. Similar to optical measurements
the radio signal of an air shower gives access to the longitudinal shower profile
and, hence, the primary shower energy and particle type and
mass~\cite{Huege:2008tn,Kalmykov:2009x1,Apel:2012re,AlvarezMuniz:2012sa,deVries:2013dia}.
Moreover it is possible to use the radio detection technique with almost 100\% duty cycle.
Therefore, with the availability of suitable electronics and improved shower simulation
methods~\cite{deVries:2011pa,AlvarezMuniz:2011bs,Huege:2013vt,Huege:2014th}, the study of radio emission
by air showers is receiving increasing
attention in recent years~\cite{Falcke:2005tc,Ardouin:2006nb,Ardouin:2010gz,Abreu:2012pi}. 

Extensive air showers consist of a disk of high-energy particles
traversing the atmosphere. With the thickness of this disk being of
the order of $1$\,m up to tens of meters from the core, charged particles emit
electromagnetic waves coherently mainly at frequencies below about $100$\,MHz,
corresponding to a wave length of $3$\,m.
However, large-scale exploitation of this signal and, in particular,
triggering on the radio pulses directly, is hampered by considerable
background radiation from natural sources and a significant amount of
transients in this frequency range due to anthropogenic sources.

In comparison, observations in the lower GHz (microwave) range would
have significant benefits. First, the background noise is extremely
low at these frequencies and the atmosphere is almost perfectly
transparent for such waves, nearly independent of cloud
coverage. Second, reliable low-noise receiver systems are commercially
available because they have been developed and refined for satellite
TV antennas for many decades.

Until recently, radio emission from air showers at GHz frequencies has not been
considered a promising detection channel as the wavelengths are much
smaller than the typical size of the shower disk.
Early measurements covered only frequencies
up to $550$\,MHz and showed a strong, exponential suppression of higher
frequencies~\cite{Fegan:1968aa,Spencer:1969x1,Fegan:1969x1,Fegan:2011fb}.
However, during the 2006/2007 flight of the balloon borne detector ANITA searching
for radio pulses originating from neutrinos interacting in the Antarctic ice
sheet, signals of air showers with frequencies up to $900$\,MHz were discovered
serendipitously~\cite{Hoover:2010qt}.

In addition, Gorham et al.~\cite{Gorham:2007af} pointed out that the numerous,
slow ionization electrons induced by the high-energy particles of the shower
disk are expected to emit molecular bremsstrahlung at GHz frequencies. Several
experiments were set up to search for such a signal using test
beams~\cite{Gorham:2007af,Williams:2013pza,AlvarezMuniz:2013sza,Bluemer:2013aa,Conti:2014dba}
and air shower
detectors~\cite{Gorham:2007af,AlvarezMuniz:2012ew,Facal:2013x1,Gaior2013x1,Smida:2011cv,Smida:2013rza}.

First unambiguous detections of microwave signals of air showers
in the $3$--$4$\,GHz range were reported recently by two of these
experiments, EASIER at the Pierre Auger Observatory~\cite{Facal:2013x1}
and CROME~\cite{Smida:2013rza}.
In this work we present the first study of the overall features of the
microwave radiation of air showers using data of the CROME experiment.
We show that the measured signal is compatible with the high-frequency tails of the
geomagnetic and Askaryan (i.e.\ coherent radiation by the charge excess
in the shower front) emission processes.


\textit{The CROME experiment} -- The CROME
experiment~\cite{Smida:2011cv,Smida:2013rza} was located within the KASCADE-Grande
air shower array~\cite{Apel:2010zz}.
It consisted of various radio antennas covering a wide
range of frequencies from about $1$\,MHz up to $12$\,GHz. The data
reported in this Letter were taken with three C~band
($3.4$--$4.2$\,GHz) antennas, each consisting of a parabolic reflector
with a diameter of $335$\,cm. The focal plane of each reflector was
equipped with a $3\times3$ matrix of linearly polarised C~band
receivers. Hence, each antenna provided nine sharp, pencil-like beams
with opening angles of $\lesssim2$\textdegree\ (half-power beamwidth)
and about $41$\,dBi gain. In the course of the experiment, the four
outermost beams of each antenna were upgraded with additional
receivers for the measurement of two polarisation directions.

The three C~band antennas were pointed into different
directions---$15$\textdegree\ from zenith towards magnetic south,
zenith, and $15$\textdegree\ from zenith towards magnetic north---to
observe parts of the sky with varying angles relative to the local geomagnetic
field ($10$\textdegree, $25$\textdegree, and $40$\textdegree,
respectively). The pointing directions and radiation patterns of the
antennas were validated using a calibrated airborne
transmitter~\cite{Smida:2011cv}. During the main operating period
between May~2011 and November~2012 the experiment was gradually
expanded to this final setup, with effective operating periods of
$551$ days for the vertical pointing, $378$ days for the antenna
pointed towards north, and $215$ days for the antenna pointed towards
south.

Logarithmic power detectors were used to measure the envelopes of the
antenna signals within an effective bandwidth of $\sim 600$\,MHz
around $3.8$\,GHz. The response time of the complete system was $\sim
3$\,ns with a total system noise temperature of $\lesssim 90$\,K.

The read-out of all receivers was triggered by the detection of
high-energy air showers with KASCADE-Grande, resulting in an effective
energy range of $3\times10^{15}$ to $10^{18}$\,eV.
The signals recorded within a window of $\pm10$\,\textmu s with respect to
the KASCADE-Grande trigger were digitised and stored for offline
analyses together with the results of the standard shower reconstruction
of KASCADE-Grande~\cite{Apel:2010zz}.


\begin{figure}[htb!]
  \begin{center}
    \includegraphics{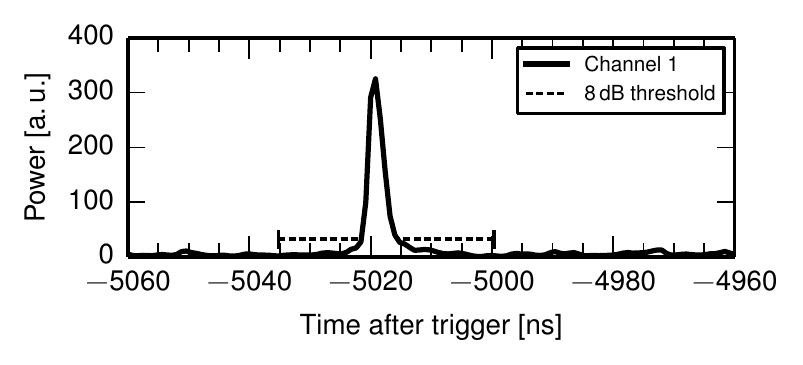}\\
    \includegraphics{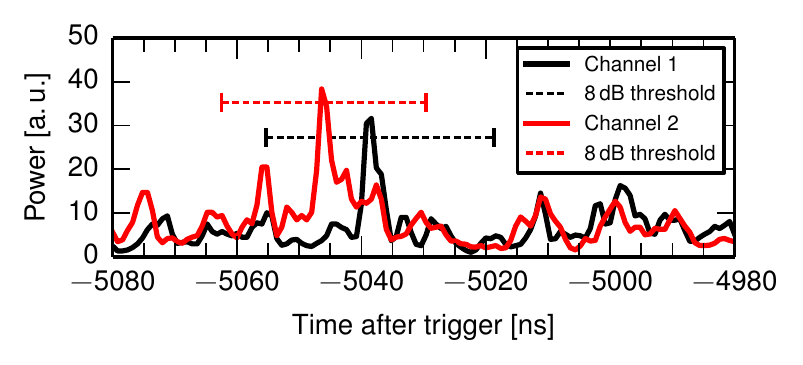}
    \caption{Power (linear scale) as a function of time relative to the
      KASCADE-Grande trigger for the event with the highest
      signal-to-noise ratio (top) and a stereo event (bottom). The
      $8$\,dB thresholds for the signal search are shown as dashed
      lines, with the horizontal extents indicating the time windows
      in which signals are expected.\label{Fig-event}}
  \end{center}
\end{figure}

\textit{Event selection} -- The common operating time of
KASCADE-Grande and CROME amounts to about $10\,000$ hours,
corresponding to about $19\,000$ showers with energies $>10^{16}$\,eV having
crossed the field of view of at least one receiver.
Less than $1$\,\% of the events were discarded due to external interference
or extreme weather conditions.

The expected arrival time of the microwave signal from each shower was
calculated using the reconstructed shower geometry, accounting for the
altitude-dependent refractive index and the measured signal
propagation times in the detectors. The typical uncertainty of the
expected signal arrival time is $\sim50$\,ns. For each trace, the signal strength
within this time window is quantified relative to the average noise
level outside of the window.

Selecting air showers with energies above $3\times10^{16}$\,eV and
signal-to-noise ratios exceeding a threshold of $8$\,dB yields $37$
event candidates with microwave signals.
The expected number of noise signals
exceeding the threshold level is estimated from data by repeating the
analysis for shifted time windows and is found to be $9.4\pm0.2$.

The time traces of microwave signals of two air showers are shown in
Fig.~\ref{Fig-event}. The top panel shows the largest signal measured
with CROME, $17.7$\,dB above the noise level. The energy of the shower
was $2.5\times10^{17}$\,eV, the zenith angle $5.6$\textdegree, and the
core distance to the antenna $120$\,m. One of the two stereo events is
shown in the bottom panel (shower energy $3.7\times10^{16}$\,eV,
zenith angle $3.7$\textdegree, core distance $110$\,m). The absolute
timing of all detected pulses is well within the expected time window
and, for the stereo events, the relative delays between the pulses are
in good agreement with the expectations (cf.\ Fig.~\ref{Fig-event}).


\textit{Properties of the microwave signal} --
The reconstructed trajectory of each selected air shower typically intersects
the fields of view of five receivers. Only the receivers
viewing the air shower at high altitudes, usually above $2$\,km,
detected a signal. Hence, the main emission region is close to the
expected maximum of the shower development ($\sim 4$\,km above ground
for a vertical shower with an energy of $10^{17}$\,eV).

\begin{figure}[htb!]
  \begin{center}
    \includegraphics{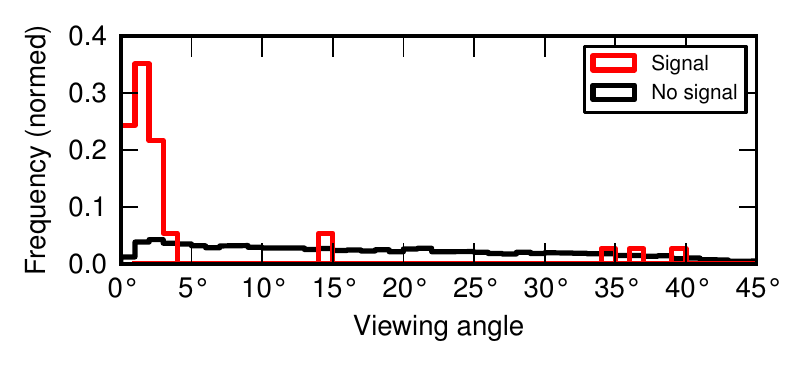}
    \caption{Distribution of viewing angles for showers passing the
      KASCADE-Grande and geometry selection criteria for the $37$
      receivers with microwave signals above the search threshold (red)
      and $\sim15000$ receivers without (black).\label{Fig-viewing-angles}}
  \end{center}
\end{figure}

In Fig.~\ref{Fig-viewing-angles}, the distribution of the viewing
angles---the angles between the shower axes and the boresight axes of
the receivers---is shown for events passing the shower selection
criteria. The distribution of the events with a microwave signal is
sharply peaked below $4$\textdegree\ and thus differs significantly
from that of the events without a signal. Hence, the majority of the
showers was detected from their forward direction. Taking into account
the fields of view of the receivers ($\sim 2$\textdegree) and the
uncertainties of the shower geometry ($\sim 1$\textdegree), the angles
of emission are compatible with the Cherenkov angle in air ($\sim
1.1$\textdegree\ at $4$\,km).

Further evidence for an emission in the forward direction is found in
the distribution of the core positions of the detected showers:
the core positions form a ring structure at a distance of $70$--$150$\,m
around the antennas.
This is compatible with the size of the Cherenkov cone projected on ground from altitudes
higher than $3$\,km.


\begin{figure}[t]
\begin{center}
\includegraphics{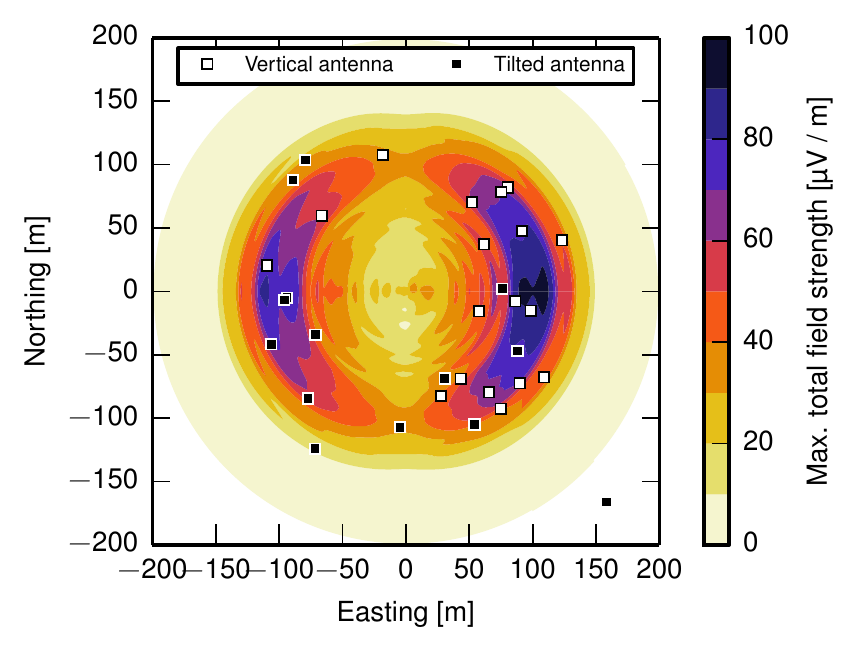}\\
\includegraphics{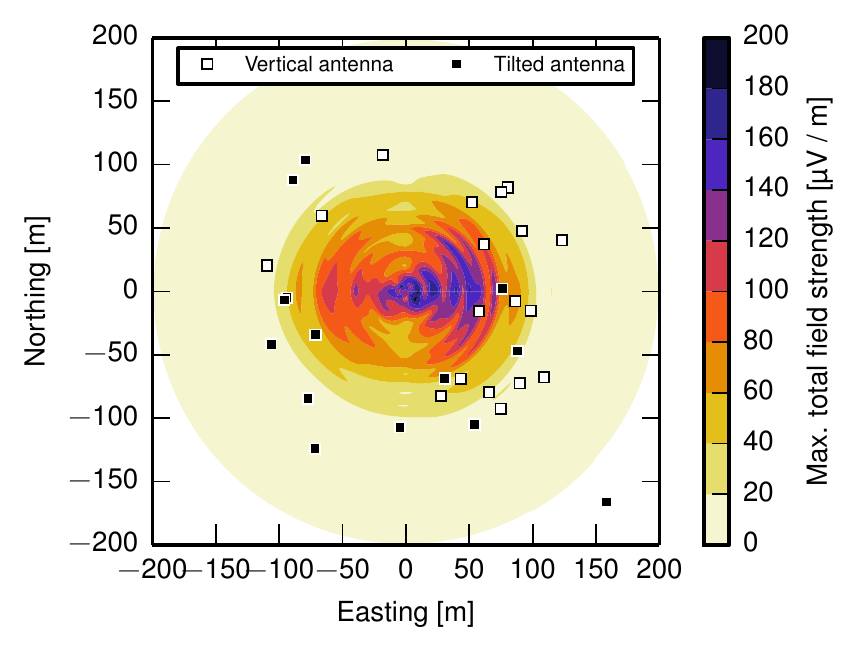}
 \caption{Positions at which a microwave signal has been detected
   relative to the shower core at $(0,0)$.
   The colour contours indicate the maximum total field
   strength at ground level predicted by CoREAS for a typical vertical
   ($X_{\textnormal{max}}=658$\,g\,cm$^{-2}$, upper panel) and a very deep
   vertical shower ($X_{\textnormal{max}}=895$\,g\,cm$^{-2}$, lower panel)
   of $10^{17}$\,eV is shown.
   \label{Fig-cores}
 }
\end{center}
\end{figure}

\textit{Interpretation} -- The observed radio signal of air showers
in the MHz range stems mainly from two emission processes~\cite{Huege:2010vm}.
Firstly, the deflection of the electrons and positrons of the shower disk in
the Earth's magnetic field results in time-varying transverse currents
(geomagnetic radiation)~\cite{Kahn:1966x1,Huege:2003up}. Secondly, the shower
disk contains $20$--$30$\,\% more electrons than positrons, leading to
a varying charge excess and, hence, electromagnetic radiation (Askaryan
effect)~\cite{Askaryan:1961x1,Askaryan:1965x1,Kahn:1966x1}.
Several simulation codes are available for describing these emission
processes in detail.
In the following we will use CoREAS~\cite{Huege:2013vt} to obtain predictions
for the expected emission features in the GHz range~\cite{James:2010vm,AlvarezMuniz:2012sa,Huege:2013vt,deVries:2013dia}.

The basic features of the expected microwave signal at ground and the
dependence on the longitudinal shower profile are illustrated in
Fig.~\ref{Fig-cores} for two vertical showers of $10^{17}$\,eV simulated
with CoREAS: a typical shower with a depth of maximum
$X_{\textnormal{max}}=658$\,g\,cm$^{-2}$ (upper panel) and
a deep proton shower with $X_{\textnormal{max}}=895$\,g\,cm$^{-2}$ (lower panel).

To compare the measured events
with the predicted radio signal we first improve the purity of the event
sample. By considering only events with a viewing angle less than
$4$\textdegree\ (cf.~Fig.~\ref{Fig-viewing-angles}) we obtain $31$ showers
(including two stereo observations)
for an expected number of $1.1\pm0.1$ noise signals.
For these events, the positions at which the GHz signal is detected
relative to the shower core are shown in Fig.~\ref{Fig-cores}.
Both the structure of the Cherenkov-like ring and the asymmetries
observed in data are qualitatively well reproduced by the typical
shower.

Considering nearly vertical showers, the superposition of the mainly east-west
polarised electric field of geomagnetic radiation and the radially inward-polarised
field due to the Askaryan effect leads to a pronounced east-west
asymmetry in the overall signal strength.
With $14$ and $3$ events detected with the vertically pointing antenna
east and west of the shower core, respectively, this asymmetry is
visible in data.
For the antenna oriented towards north, no
  significant east-west asymmetry is observed ($4$ vs.\ $8$ events) as
  can be expected for an increasing dominance of geomagnetic emission
  due to the larger geomagnetic angle.

Thus, the GHz observations are consistent with the known
signatures of the radio emission of high-energy charged particles in
an air shower which is collimated in a cone about the Cherenkov angle
due to the refractive index
of air, an effect first described by de Vries et al.~\cite{deVries:2011pa}.
An antenna located on or near this cone---projected from the main emission
region near the shower maximum on ground---receives a radio pulse of only
a few ns duration. Hence, near the emission cone a harder frequency
spectrum and coherence up to GHz frequencies are expected.

A detailed comparison of the observed signal amplitudes with the CoREAS
predictions can only be done after a full end-to-end calibration of CROME
and is beyond the scope of this article.

\begin{figure}[htb!]
\begin{center}
\includegraphics{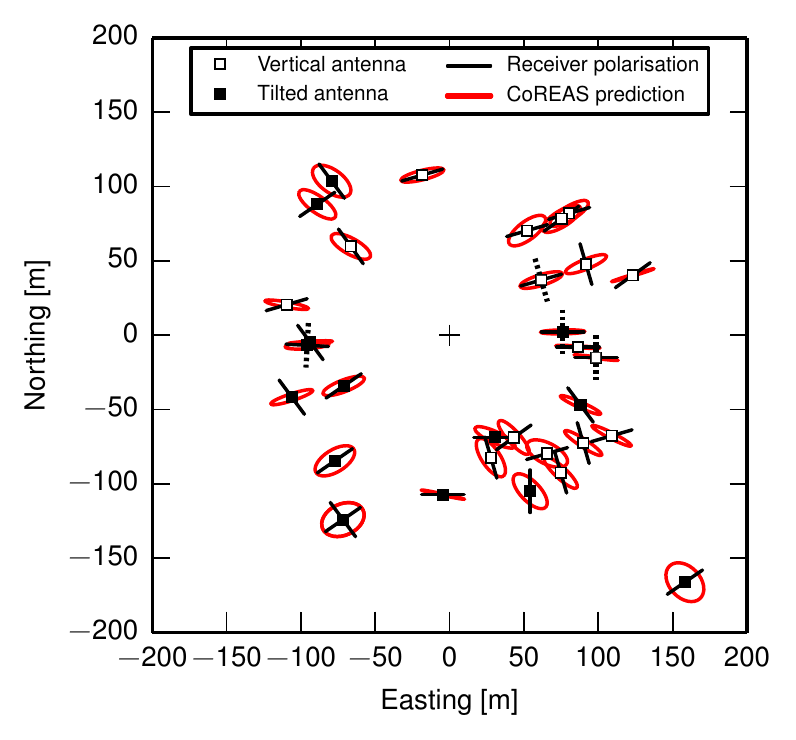}
 \caption{Polarisation directions of receivers in which a microwave
   signal was detected (black lines). In addition, the predicted
   polarisation ellipses simulated with CoREAS for iron showers are
   shown in red. For dual-polarised receivers, the polarisation
   direction in which no signal was detected is shown as dashed
   line. The shower core is always at $(0,0)$.
   \label{Fig-polarisation}
 }
\end{center}
\end{figure}

To compare the measured polarisation directions with CoREAS
predictions we simulated showers for each observed event (cf.~Fig.~\ref{Fig-polarisation}).
Motivated by the composition measurement of KASCADE-Grande~\cite{Apel:2011mi} iron nuclei were used.
The polarisation information can be compared directly
for the three events detected with dual-polarised receivers.
It is found that the signal was always detected only with the
receiver whose polarisation direction was close to that predicted
in the simulations, but the statistics of such events is too small to
draw conclusions.

Therefore, we applied a detector simulation to the time traces
obtained from the CoREAS simulations and calculated the loss in
detectable power due to the projection of the predicted electric field
vector onto the polarisation direction of the receivers (polarisation
loss). Assuming that the time-dependent local electric field vector is
correctly described by the CoREAS predictions, we find an average
polarisation loss of $34$\,\% for the measured
microwave signals. For unpolarised pulses from incoherent radiation
with a flat frequency spectrum, an average polarisation loss of
$50.0\,\% \pm 3.5\,\%$ would be expected.
Therefore, within this model, the hypothesis that
  the observed radio emission is unpolarised is rejected with a
  significance of $4.7\,\sigma$.

The polarisation pattern of the signal disfavours an explanation
in terms of molecular bremsstrahlung as dominant emission mechanism
which is in qualitative agreement with recent
findings~\cite{Williams:2013pza,AlvarezMuniz:2013sza,Conti:2014dba}.

\textit{Conclusions and Outlook} -- 
Using air showers measured with CROME in coincidence with KASCADE-Grande we
have determined fundamental properties of the microwave emission of air
showers in the forward direction. 
We have shown that the spatial and angular distributions of the microwave
signal are in good agreement with the extension of the well-known
radio emission processes at tens of MHz into the GHz range
close to the Cherenkov
angle~\cite{deVries:2011pa,AlvarezMuniz:2012sa,Huege:2013vt}.
The collected polarisation information strongly supports this conclusion.

We have illustrated that this technique can be successfully used for
the measurement of extensive air showers. The main advantages for
the observation in the GHz range are the low background noise, the nearly
perfect transparency of the atmosphere at microwave frequencies, as well
as the availability of a well-developed technology for microwave
detection~\cite{Gorham:2007af}.

One can envisage various applications of measuring air showers with setups
similar to the one presented in this Letter. For example, the data collected
by the balloon-borne ANITA detector at the South Pole~\cite{Hoover:2010qt}
or future balloon or satellite experiments of this
type~\cite{Gorham:2011mt,Romero-Wolf:2013etm} could be experimentally
verified and calibrated.

Additionally, in contrast to optical detectors such as imaging atmospheric
Cherenkov telescopes~\cite{Voelk:2008fw}, the possibility of measuring
with a nearly $100$\,\% duty cycle and using simple metallic
reflectors instead of optical mirrors could make this measurement
technique promising for air showers with energies above hundreds of TeV.
Particularly a measurement of inclined air showers, where the footprint of
the microwave signal extends over hundreds of meters, could compete with
optical detectors.


\textit{Acknowledgments} -- It is our pleasure to acknowledge the
interaction and collaboration with many colleagues from the Pierre
Auger Collaboration, in particular Peter Gorham, Antoine
Letessier-Selvon and Paolo Privitera. This work has
been supported in part by the KIT start-up grant 2066995641, the
ASPERA project BMBF 05A11VKA and 05A11PXA, the Helmholtz-University
Young Investigators Group VH-NG-413 and the National Centre for
Research and Development in Poland (NCBiR) grant ERA-NET-ASPERA/01/11.


\end{document}